\documentclass[aps,prd,10pt,superscriptaddress,nofootinbib,showkeys,twocolumn]{revtex4-2}

\usepackage[T1]{fontenc}
\usepackage{lmodern}
\usepackage[scaled=1]{XCharter}
\usepackage[scaled=.95]{beramono}
\usepackage[charter,ncf,scaled=1.05]{newtxmath}
\usepackage[bb=dsserif,bbscaled=1.15]{mathalpha}

\usepackage{mathtools,siunitx,physics}
\usepackage{graphicx}
\usepackage[normalem]{ulem}
\usepackage{enumerate}
\usepackage{silence}
\ExplSyntaxOn
\msg_redirect_name:nnn{siunitx}{physics-pkg}{none}
\ExplSyntaxOff
\usepackage[dvipsnames]{xcolor}
\definecolor{theme}{rgb}{0.7, 0, 0.35}
\usepackage[colorlinks, hypertexnames=false, allcolors=theme, bookmarksnumbered=true]{hyperref}
\usepackage[capitalise]{cleveref} 
\usepackage{orcidlink}
\usepackage{xspace}

\newcommand{\etal}{\textit{et~al.}\@\xspace}
\newcommand{\ie}{i.e.,\@\xspace}

\crefname{section}{Sec.}{Secs.}
\Crefname{section}{Section}{Sections}
\crefname{appendix}{App.}{Apps.}
\Crefname{appendix}{Appendix}{Appendices}
\newcommand{\Cite}[1]{Ref.~\cite{#1}}
\newcommand{\Cites}[1]{Refs.~\cite{#1}}

\DeclareMathAlphabet{\mathpzc}{OT1}{pzc}{m}{it}
\newcommand{\mpz}[1]{\mathpzc{#1}}

\newcommand{\msf}[1]{\mathsf{#1}}

\newcommand{\mk}{\mspace{1mu}}
\renewcommand{\vec}[1]{\boldsymbol{#1}}

\newcommand{\pd}{\partial}

\newcommand{\del}{\vec{\nabla}}

\newcommand{\sub}[1]{_\mathrm{#1}}

\providecommand{\vv}{}
\renewcommand{\vv}{\vec{v}}
\newcommand{\ee}{\vec{E}}
\newcommand{\bb}{\vec{B}}
\newcommand{\jj}{\vec{J}}

\newcommand{\cm}{C_5}
\newcommand{\cw}{C_\omega}
\newcommand{\kk}{k_\mu}
\newcommand{\cf}{C_{\rm flow}}
\newcommand{\hh}{\mpz{h}}
\newcommand{\ttt}{\tau}

\newcommand{\sect}[1]{\medskip\noindent\textit{#1.}---}

\begin{document}

\title{A charge-flow instability in plasmas with charge fluctuations}

\author{Deepen~Garg\,\orcidlink{0000-0001-5226-1913}}
\email{dgarg@uni-bonn.de}
\affiliation{Argelander-Institut für Astronomie, Universität Bonn, Bonn 53121, Germany}

\author{Jennifer~Schober\,\orcidlink{0000-0001-7888-6671}}
\email{schober@uni-bonn.de}
\affiliation{Argelander-Institut für Astronomie, Universität Bonn, Bonn 53121, Germany}

\date{\today}


\begin{abstract}
We present a linear instability in magnetized plasmas with charge fluctuations. It is driven by an electric current proportional to the charge chemical potential $\mu$ and the bulk velocity.
This charge-flow (C-flow) instability has hitherto not been considered in standard magnetohydrodynamics or its chiral extensions.
We derive its dispersion relation and maximum growth rate, and confirm them with direct numerical simulations. We also show that the C-flow instability persists even for zero-mean fluctuations of $\mu$, vanishing resistivity, and a magnetic Prandtl number of unity.
\end{abstract}

\maketitle

\sect{Introduction}
The chiral anomaly, the quantum nonconservation of the axial current of
massless Dirac fermions in the presence of gauge fields
\cite{adler:69,Bell:1969ts}, survives at macroscopic scales owing to its topological nature. At these scales, it can give rise to anomalous transport, most notably the chiral magnetic effect (CME),
\(\propto \mu_5 \bb\) \cite{ref:vilenkin80,ref:fukushima08,ref:rybalka19}, and the
chiral vortical effect (CVE),
\(\propto \mu \mu_5 \vec\omega\) \cite{ref:rybalka19,ref:vilenkin79,ref:son09}, where
\(\vec\omega\) is the vorticity of the fluid
(see \Cite{ref:kharzeev16} for a review). Here, $\mu_5$ is the chiral
chemical potential quantifying the asymmetry between left- and
right-handed fermions, and $\mu$ is the chemical potential quantifying
the asymmetry between fermions and antifermions. Both chemical potentials are normalized by the fluid temperature. For charged fermions,
the CME current enters the classical magnetohydrodynamics (MHD)
equations and drives the chiral plasma instability
\cite{ref:joyce97,Boyarsky:12,ref:rogachevskii17,ref:schober24prl,ref:schober24prd},
with applications ranging from condensed matter physics \cite{ref:son13,ref:huang15,ref:xiong15,ref:li16} to heavy-ion collisions \cite{kharzeev:24}, to proto-neutron stars \cite{tex:ohnishi14,ref:yamamoto16,ref:masada18,ref:matsumoto22,tex:dehman24,ref:dehman25,tex:dehman26,ref:sigl16,ref:grabowska15}, to
early-Universe magnetogenesis and baryogenesis (see \Cite{Kamada:23} for a review of the astrophysical and cosmological applications). Importantly, processes that generate $\mu_5$ typically generate $\mu$ as well.

Here, we present another instability that arises when $\mu \neq 0$. As we show below, it does not require \(\mu_5\), and is thus not strictly a chiral phenomenon even though we derive the governing equations in the context of chiral MHD in our companion paper \cite{my:chiralmhd_theory}. This instability is driven by an extra current \(\propto \mu \vv\) for charged fermions that is absent in classical MHD, where $\vv$ is the plasma bulk velocity. Notably, fluctuations of $\mu$ with zero volume-averaged mean are sufficient to trigger it. We call it the charge-flow (C-flow) instability.

This additional C-flow term is necessary whenever there is a deviation from quasineutrality, \ie nonzero $\mu$ as defined above. However, to our knowledge, it has not been studied in the existing chiral MHD literature with nonzero $\mu$ \cite{ref:schober24prd,ref:schober24prl,ref:huang18,ref:wang24,DAS:25,tex:wang25,wang:26}. In fact, this term can exceed the CVE current \cite{ref:wang24,DAS:25,tex:wang25,wang:26} (see the appendix) and can even be comparable to the CME current in certain scenarios (see below and \cite{my:chiralmhd_theory}), necessitating its inclusion in a comprehensive analysis. In the plasma physics literature, the standard treatments assume either rapid charge screening \cite{ref:komissarov07,ref:palenzuela09} or force-free conditions \cite{ref:goldreich69,ref:contopoulos99,ref:spitkovsky06}. The C-flow instability, distinct from electrostatic charge-flow instabilities such as the diocotron instability \cite{ref:levy65}, arises in the regime where neither simplification holds (see the discussion of saturation below).

The C-flow instability can be understood to be in the same class as other mechanisms which have an electromotive force (EMF) proportional to the bulk velocity. These include the Yoshizawa cross-helicity dynamo \cite{ref:yoshizawa90,ref:yokoi13} and the $\expval{\vv \cdot\jj}$ effect of \Cite{ref:radler10}. There, the EMF arises as a turbulent transport coefficient, and vanishes at unit magnetic Prandtl number, the ratio of kinematic viscosity to magnetic diffusivity. The C-flow term is physically distinct: it requires no turbulence, and operates at unit magnetic Prandtl number, where these turbulent EMFs vanish.

\sect{Theoretical setup}
We consider the one-fluid approximation for a plasma consisting of charged fermions. Although the C-flow instability requires no chiral effects, we keep the discussion general with a view to applications in early-Universe chiral MHD. We assume small chemical potentials and subrelativistic $\vv$,
\begin{gather}
    \mu^2, \mu_5^2, v^2 \ll 1 \,,
    \qquad v = \abs{\vec v} \,.
    \label{eq:smallmu}
\end{gather}
The full chiral-MHD equations are derived in our companion paper \cite[Eqs.~(39)-(44)]{my:chiralmhd_theory}. Following its conventions, we set \(c = k\sub{B} = 1\) while keeping \(\hbar\) and the elementary charge \(e\) explicit. In the appendix, we reduce the equations to isolate the C-flow instability, yielding
\begin{align}
    \pd_\ttt \mu
    &= - \del \cdot \qty(\mu \vv + \cm \mu_5 \bb') - D_\mu \del^4 \mu \,,
    \label{eq:mu}
    \\[2pt]
    \pd_\ttt \mu_5
    &= - \del \cdot \qty(\mu_5 \vv + \cm \mu \bb') - D_5 \del^4 \mu_5
    + \lambda_5 \mk\mk \ee' \cdot \bb',
    \label{eq:mu5}
    \\[2pt]
    \pd_\ttt \bb'
    &= - \del \times \ee'\,,
    \label{eq:ind}
    \\[2pt]
    \ee' &= - \vv \times \bb'\mkern-1mu
    + \eta \left( \del \times \bb'\mkern-1mu - \kk \mu_5 \bb'\mkern-1mu
    - \cf \kk \mu \vv \right),
    \label{eq:elec}
    \\[2pt]
    {\frac{D \vec{v}}{D \ttt}}
    &= \frac{\rho_0}{\rho} (\vec{\del} \times \vec{B}') \times \vec{B}'
    - \frac{\vec{\del} p}{\rho} + \frac{\vec{\del} \cdot (2 \nu \rho \vec{\msf{S}})}{\rho} \,,
    \label{eq:v0}
    \\[2pt]
    \frac{D\rho}{D \ttt}
    &= - \rho \vec{\del}  \cdot \vec{v} \,,
    \label{eq:mass0}
\end{align}
where $\lambda_5 = 1.97 \mk \kk$, $\bb' \equiv \bb/\sqrt{\mu_0\rho_0}$, $\ee' \equiv \ee/\sqrt{\mu_0\rho_0}$, \(\rho_0\) is the background plasma density, \(\mu_0\) is the vacuum permeability, $\ttt$ is the comoving time, $D/D\ttt \equiv \pd_\ttt + \vv\cdot\del$, $\vec{\msf S}$ is the rate-of-strain tensor, and $\nu$ is the kinematic viscosity. Here, \(\kk, \cf, \cm, D_\mu\) and $D_5$ are constants defined in the appendix with further details in \Cite{my:chiralmhd_theory}.

The C-flow instability arises from the last term in \cref{eq:elec}, which can exceed the preceding CME term when \(\mu \approx \mu_5\) and the magnetic field and the velocity field are in equipartition. In the absence of the C-flow term, the system reduces to standard chiral MHD leading to the chiral plasma instability, which saturates through the conservation of total chirality plus magnetic helicity \cite{my:chiralmhd_theory}, \(\expval{\mu_5} + 0.98 \mk \kk \expval{\vec{A}' \cdot \bb'} = {\rm const.}\),
with $\bb' \equiv \del\times\vec A'$. As $\mu_5$ is converted into magnetic helicity, the chiral dynamo growth rate $\propto\mu_5^2$ falls and the dynamo saturates. The C-flow term has no such reservoir, leading to a runaway. We return to this below, in the discussion of saturation.

\sect{Charge-flow battery}
In the simplest case of zero magnetic field but nonzero $\mu$ and velocity perturbations, the C-flow EMF acts as a battery,
\begin{gather}
    \pd_\ttt \bb' = \del \times \qty(\eta \cf \kk \mu \vv) \,.
\end{gather}
This is consistent with the usual understanding that charge separation can drive electric currents and, consequently, magnetic fields. The C-flow term, however, not only produces the initial field, but also leads to an instability, which affects the magnetic evolution and associated processes such as the chiral dynamo.

\sect{Linear charge-flow instability}
We first establish that the C-flow term renders the system linearly unstable and derive the growth rate. We linearize around a static homogeneous background magnetic field, $\bb' = \vec B_0 + \vec b$, with no background velocity. Both \(\mu_5\) and \(\mu\) are kept constant with \(\mu_5\) assumed small, resembling the state after the chiral dynamo saturates. We stress that $\mu_5$ is not required for the C-flow instability (\(\mu_5\) is in fact set to zero in some runs below). Let us also assume an isentropic equation of state, \(\pd p = c_s^2 \pd \rho\), where \(c_s\) is the sound speed. With this, the reduced linear equations become
\begin{gather}
    \pd_\ttt \vec b
    = \del \times \qty(\vv \times \bb_0
    - \eta \del \times \vec b
    + \cf \eta \kk \mu \vv) \,,
    \\
    \pd_\ttt \vec v
    = (\vec{\del} \times \vec{b}) \times \vec{B}_0
    - c_s^2 \frac{\del \rho}{\rho}
    + \nu \qty(\del^2 \vv + \frac{1}{3} \del \del \cdot \vv) \,,
    \\
    \frac{1}{\rho} \pd_\ttt \rho = - \del \cdot \vv \,,
\end{gather}
from which we obtain the following wave equation
\begin{multline}
    \qty[\omega \qty(\omega + i \nu k^2) - \qty(\bb_0 \cdot \vec k)^2 + i \eta \qty(\omega + i \nu k^2) k^2] \vec b
    \\
    = \frac{\omega \qty(\omega + i \nu k^2) k^2 \qty(\vec \Pi_k \cdot \bb_0)}{\omega^2 - k^2 c_s^2 + 4 i \nu k^2 \omega/3} \qty(\vec b \cdot \bb_0)
    \\
    + \qty[\cf \eta \kk \mu \qty(\bb_0 \cdot \vec k)
    - \eta \kk \mu_5 \qty(\omega + i \nu k^2)] \vec k \times \vec b \,,
    \label{eq:wave}
\end{multline}
where \(\vec \Pi_k \equiv \mathbb{1} - \vec k \vec k/k^2\) is the projector onto the plane perpendicular to $\vec k$, $\mathbb{1}$ is the identity matrix, and \(\omega\) is the frequency of a monochromatic wave \(\vec b \propto \exp(i \vec k \cdot \vec x - i \omega t)\). For perpendicular propagation, \(\vec k \perp \bb_0\), the first term on the right-hand side of \cref{eq:wave} yields magnetosonic waves alongside the chiral dynamo and resistive and viscous decay. More importantly, it does not couple to the C-flow term, which is proportional to \(\bb_0 \cdot \vec k\) and thus drives a parallel mode. We therefore consider parallel propagation, \(\bb_0 \cdot \vec k = k \mk B_0\), for which the wave equation simplifies significantly
\begin{multline}
    \qty(\omega\qty(\omega + i \nu k^2) - k^2 B_0^2 + i \eta \qty(\omega + i \nu k^2) k^2) \mk \vec b
    \\
    + \eta \mk \kk \qty(\mu_5 \qty(\omega + i \nu k^2) - \cf \mu B_0 k) \qty(\vec k \times \vec b) = 0 \,,
\end{multline}
from which we obtain the eigenmodes, the two magnetic helicities, both of which satisfy \(b^2 = 0\),\footnote{Note that \(\vec b\) denotes the complex Fourier amplitude rather than a real-valued field in position space.} and the dispersion relation
\begin{multline}
    \omega(k) = -i \frac{(\eta+ \nu) k^2 + \hh \eta \mu_5 \kk k}{2}
    \\
    \pm i k \sqrt{ \frac{\qty(\eta k - \nu k + \hh \eta \mu_5 \kk)^2}{4}
    - B_0^2
    - i \hh \eta \kk \cf \mu B_0 } \,,
    \label{eq:disp1}
\end{multline}
where \(\hh = \pm 1\) labels the two helicities.

\Cref{eq:disp1} reveals several features of the C-flow instability. For $\cf = 0$, we recover the usual chiral dynamo with resistive decay and Alfv\'en-like oscillations. The C-flow coupling enters the square root as an imaginary contribution, the signature of a non-Hermitian linear operator, and renders the system unstable. Without the chiral term, the two helicities grow at the same rate, differing only in the sign of the oscillatory part ($\Re \omega$), in line with the parity-invariance of the C-flow instability. The instability requires a finite background field $B_0$ to operate, making it a linear magnetized plasma instability. This field could potentially be supplied by, for example, the C-flow battery itself or the chiral dynamo.

In order to obtain the characteristic length scale and time scale, let us simplify \cref{eq:disp1} by assuming \(\mu_5 = 0\) and \(\nu = \eta\), matching our simulations below. This yields the imaginary part $\omega_i$ of the growing branch to be
\begin{gather}
    \omega_i(k) = - \eta k^2
    + k \Re \sqrt{- B_0^2
    - i \hh \eta \kk \cf \mu B_0 } \,,
    \label{eq:disp2}
\end{gather}
from which we set \(\pd_k \omega_i = 0\) at $k = k\sub{max}$ to obtain
\begin{gather}
    k\sub{max} = \frac{\Re \sqrt{-B_0^2 - i \hh \eta \kk \cf \mu B_0}}{2\eta} \,,
    \\
    \abs{\omega_i}\sub{max} = \eta k\sub{max}^2 \,.
    \label{eq:gammamax1}
\end{gather}
For sufficiently small $B_0$, we obtain
\begin{gather}
    \abs{\omega_i}\sub{max} = \frac{\kk \cf \abs{\mu} B_0}{8} \,,
    \label{eq:gammamax2}
\end{gather}
which becomes independent of $\eta$ even though nonzero $\eta$ is required for the C-flow term to enter the induction equation. 
Although the dependence of $\abs{\omega_i}\sub{max}$ on the parameters is more complicated \eqref{eq:gammamax1}, it is still expected to scale almost linearly with the factors in \cref{eq:gammamax2}. We test this and quantify the deviation next.

\sect{Numerical simulations: Setup}
The full chiral-dynamo scenario is strongly nonlinear, and the linear growth rate \eqref{eq:gammamax1} applies there only transiently. To isolate the C-flow instability, we therefore designed runs without chiral effects, listed in \cref{tab:runs} (see \Cite{my:cflow_data} for run data). In all runs, the magnetic and velocity fields are initialized as Gaussian random fields with a small amplitude, and \(\mu_5 = 0\) is imposed, eliminating the CME term. Furthermore, $\mu$ is initialized as a sine wave with amplitude $\mu_{\rm ampl}$ and wavenumber $k_0$. We perform the simulations with the \texttt{Pencil Code} \cite{ref:pencil21}, using revision \texttt{5d7c032}, to solve \cref{eq:v0,eq:mass0,eq:mu,eq:mu5,eq:ind,eq:elec}. Since this instability does not require turbulent dynamics, we are able to resolve it clearly at a moderate resolution of $512^3$. The magnetic diffusivity and viscosity are \(\eta = \nu = \num{5e-2}\) in all the runs. The hyperdiffusivity for $\mu$ is chosen such that, at the Nyquist wavenumber, the diffusion rate of \(\mu\) equals the magnetic dissipation rate, yielding \(D_\mu = \num{7.5e-7}\).

\begin{figure}
    \centering
    \includegraphics[width=\linewidth]{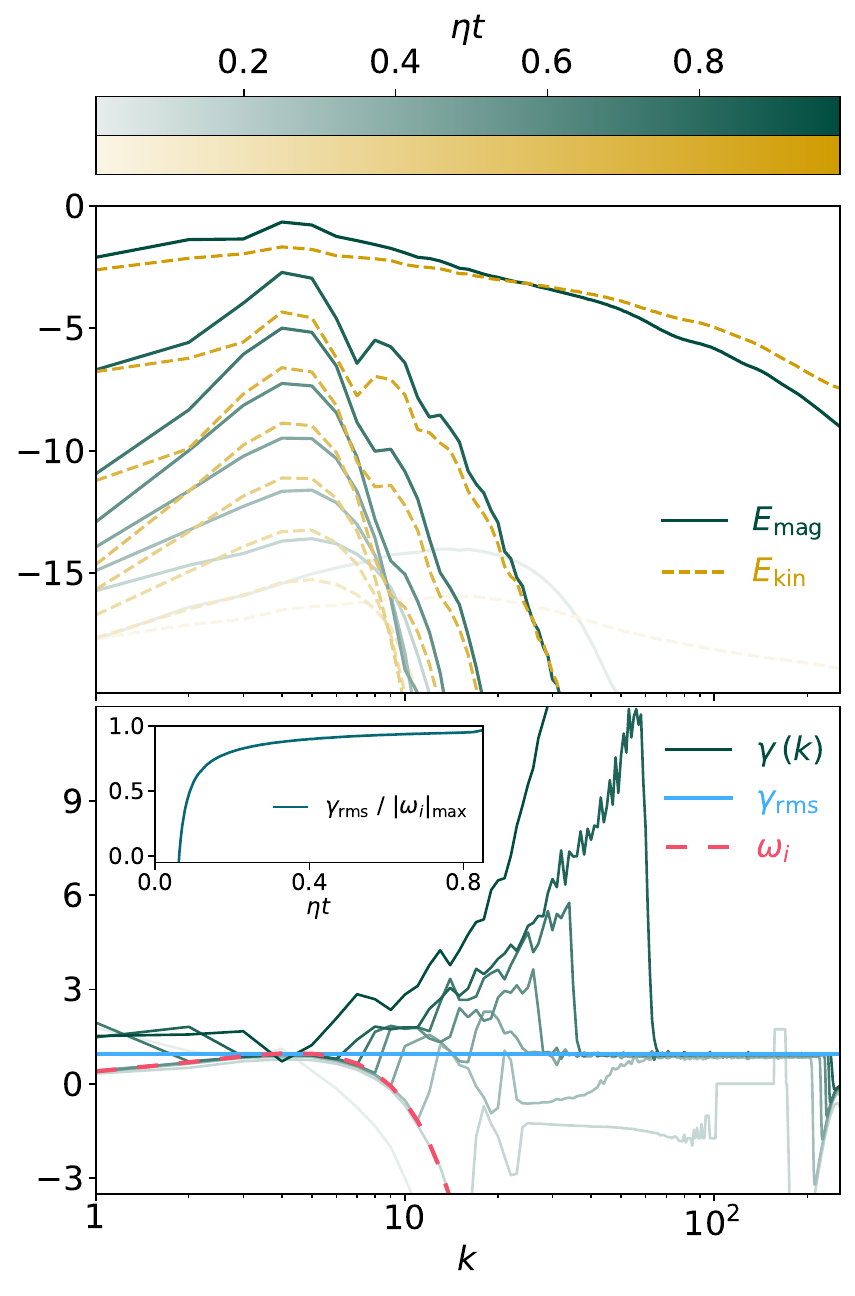}
    \caption{Spectral evolution of Run~\texttt{A0}. \textit{top}: spectra of the magnetic field (solid green) and the velocity (dashed yellow) at successive times.
    \textit{bottom}: the measured growth rates $\gamma(k)$ (solid green) and $\gamma_{\rm rms}$ (solid blue) at the same times, and the linear prediction \eqref{eq:disp2} (dashed pink). \textit{inset}: the ratio of $\gamma_{\rm rms}$ to $\abs{\omega_i}_{\rm max}$ of \cref{eq:gammamax2} over time, which approaches unity through the linear stage.
    }
    \label{fig:spec}
\end{figure}

\begin{figure}
    \centering
    \includegraphics[width=\linewidth]{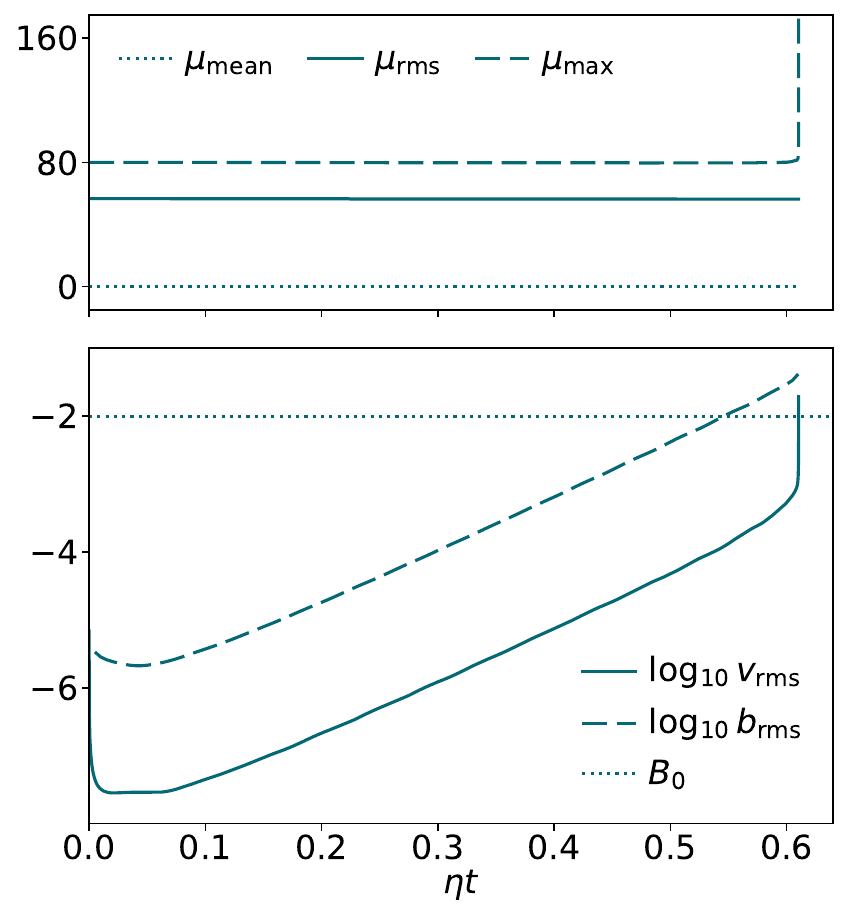}
    \caption{Time evolution of characteristic quantities for Run~\texttt{M}. \textit{top}: The spatial mean (solid), RMS (dotted), and maximum (dashed) of the chemical potential. \textit{bottom}: The RMS value of the velocity (solid) and magnetic field (dashed). }
    \label{fig:muSevol}
\end{figure}

\sect{Confirmation of linear results}
To test the scaling of \cref{eq:gammamax2}, Runs~\texttt{A0}--\texttt{A16} consider homogeneous $\mu$, \ie \(k_0 = 0\), for which $\mu$ remains constant by \cref{eq:mu}.
The fields first undergo a decline\footnote{The decline stems from the equal seeding of the growing and decaying eigenmodes by generic initial conditions (cf.~\cref{fig:muSevol} for Run~\texttt{M}).}
whose depth increases with $k$, as seen for Run~\texttt{A0} in \cref{fig:spec}.
The measured growth rate $\gamma(k)$ therefore turns positive earlier at smaller $k$, where it matches \cref{eq:disp2} (bottom panel).
The dominant mode settles at $k\sub{max}$ and grows at a constant rate throughout the run.
Higher-$k$ shells are soon dominated by overtones of this mode and grow faster than the linear prediction \eqref{eq:disp2}, revealing nonlinear coupling well before the runaway.
Despite their faster growth, these modes carry negligible energy (top panel), so the rms magnetic and velocity fields grow at the rate of the dominant mode during the linear stage. We measure this rate, $\gamma_{\rm rms}$, over the middle third (in time steps) of each simulation, where neither the initial relaxation nor the final nonlinear evolution plays a significant role.
The resulting $\gamma_{\rm rms}$ is in close agreement with $\abs{\omega_i}_{\rm max}$ of \cref{eq:gammamax2}, differing by only a few percent (bottom-panel inset).

\begin{figure}
    \centering
    \includegraphics[width=\linewidth]{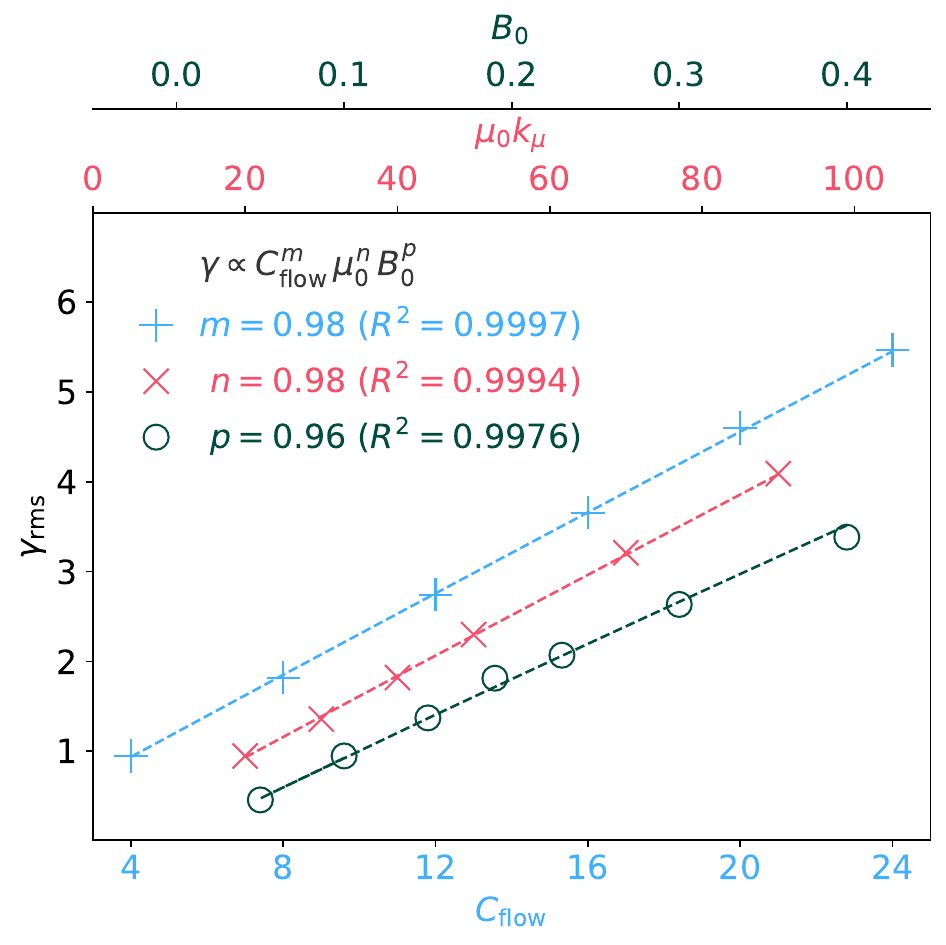}
    \caption{Measured growth rate $\gamma_{\rm rms}$ for Runs~\texttt{A0}--\texttt{A16} as a function of $C_{\rm flow}$, $B_0$ and $\mu$. The dashed line through each set of observations is the best linear fit of $\ln \gamma_{\rm rms}$ with the natural log of the three parameters. The legend shows the corresponding slope and $R^2$ of the fit.}
    \label{fig:gamma}
\end{figure}

Moreover, we find that $\gamma_{\rm rms}$ scales linearly with the parameters in \cref{eq:gammamax2}.
This is tested by considering Run~\texttt{A0} and varying each parameter individually in Runs~\texttt{A1}--\texttt{A16}. We then fit these growth rates to a power law, \ie we fit $\ln \gamma_{\rm rms}$ linearly against the natural logarithm of the varied parameter. The results are shown in \cref{fig:gamma}. We observe a nearly linear relationship for all three parameters, with high $R^2$ values, confirming a close fit to the predicted parameter dependence \eqref{eq:gammamax2}. Thus \cref{eq:disp2} captures the linear stage in full: growth rate,
dominant wavenumber, and parameter scaling.
\begin{table}
    \begin{ruledtabular}
    \begin{tabular}{*{6}{c}}
        Runs & $\cf$ & $B_0$ & $\mu_{\rm ampl} k_\mu$ & $k_0$ & $c_s$  \\
        \colrule \noalign{\vspace{4pt}}
        \texttt{M} & $26.8$ & $10^{-2}$ & $80$ & $5$ & $1/\sqrt{3}$
        \\ \noalign{\vspace{4pt}}
        \texttt{\textbf{A0}}, \texttt{A1}--\texttt{A5} 
        & $\mathbf{4},8-24$
        & $0.1$ & $20$ & $0$ & 1 
        \\ \noalign{\vspace{4pt}}
        \texttt{A6}--\texttt{A11} & $4$ 
        & $0.05-0.4$
        & $20$ & $0$ & 1 
        \\ \noalign{\vspace{4pt}}
        \texttt{A12}--\texttt{A16} & $4$ & $0.1$
        & $30-90$
        & $0$ & 1 
    \end{tabular}
    \end{ruledtabular}
    \caption{Families of simulation runs without chiral effects. See \Cite{my:cflow_data} for run data. For \texttt{A1}--\texttt{A16} runs, $\cf, B_0$ and $\mu_{\rm ampl}$ are scanned through their respective ranges one at a time, shown as markers in \cref{fig:gamma}, while the other values are kept the same as in Run~\texttt{A0}.
    }
    \label{tab:runs}
\end{table}

\sect{Evolution of a zero-mean chemical potential}
We now relax the homogeneity of the initial $\mu$, which activates its dynamical evolution, \ie \(\pd_\ttt \mu \neq 0\) in \cref{eq:mu}. Specifically, we initialize zero-mean, long-range fluctuations of $\mu$ as a sine wave with wavenumber \(k_0 = 5\). This is denoted as Run~\texttt{M} in \cref{tab:runs}.

We observe that the instability proceeds as before, with $\mu$ nearly unaffected during the linear stage; its mean, rms, and maximum values remain nearly constant. The time evolution of the relevant fields is shown in \cref{fig:muSevol}.
Eventually, both the velocity and the magnetic field blow up. This is the runaway of the non-saturating nonlinear regime mentioned earlier, to which we turn next. Note that only the maximum of $\mu$ grows appreciably in this phase, suggesting localized growth of $\mu$. The mean, however, remains constant as expected from \cref{eq:mu}.
%

\sect{Nonlinear evolution}
Once the amplified magnetic field becomes comparable to the background field, the dynamics become fully nonlinear. All the runs where the instability is active, including Run~\texttt{M} (\cref{fig:muSevol}), end in a runaway. We have verified this at higher spatial resolution (finer grid) and temporal resolution (smaller Courant number); the runaway persists. It can be understood from the structure of the governing equations.

With $\mu$ constant, the C-flow term $\propto \cf\kk\mu\vv$ is linear in $\vv$ with a constant coefficient, but the quadratic Lorentz force in \cref{eq:v0} is nonlinear. Schematically, the induction equation gives $\pd_\ttt B \sim c_1 v$ with $c_1 \propto \eta\cf\kk\mu$, while the Lorentz force gives $\pd_\ttt v \sim c_2 B^2$, so that \(\pd_\ttt^2 B \sim c_1 c_2 B^2\),
a loop in which a linear EMF feeds on the field through the quadratic Lorentz response. If this schematic picture holds, it implies a finite-time singularity; we indeed observe such a blowup in all our simulations, consistent with this picture. The observed blowup sets in shortly after the perturbations become comparable to $B_0$.

We defer a quantitative analysis of this nonlinear regime to future work, including the parametric dependence of the blowup and its possible mitigation; the latter is discussed next.

\sect{Absence of saturation}
The model considered in this work provides no saturation, even when $\mu$ evolves according to \cref{eq:mu}. This can be contrasted with the chiral dynamo, which saturates as $\mu_5$ is depleted [see the paragraph after \cref{eq:elec}]. The C-flow instability, on the other hand, draws on $\mu$, which possesses no analogous conservation law to deplete it.

This lack of a depletion mechanism is due to the MHD framework assumed in \Cite{my:chiralmhd_theory}, where we drop the displacement current and close the electric field \eqref{eq:elec} algebraically rather than dynamically. As we noted in the linear analysis, the instability exists for parallel propagation, where longitudinal charge accumulation
is magnetically unhindered. The longitudinal electric field, if treated dynamically, will therefore couple directly to the instability, with the electrostatic energy of charge separation constraining the energy reservoir.

In certain scenarios, this dynamical treatment of the electric field can be bypassed by simply considering ohmic current in the charge conservation equation and dropping the \(\vv \times \bb\) term. This leads to \(\vec J_{\rm ohm} \propto \ee/\eta\), which, via Poisson's equation, yields a resistive drag term $-\mu/\eta$. This closure using the ohmic current is, however, an approximation which is valid only when the inertia of electrons is negligible and the instability timescale is shorter than the resistive one.
It requires modification when the Langmuir oscillations are underdamped, for example in the case of the early Universe \cite{ref:baym97,ref:lebellac96, ref:blaizot96,ref:arnold00}. Charge fluctuations then decay on a much longer timescale than the plasma oscillations. A proper treatment requires a two-fluid description, which is beyond the scope of this work and is deferred to a future study. Notably, a generic process sourcing both $\mu$ and $\mu_5$ \cite{tex:gurgenidze25} would require further modification of the analysis.

While a dynamical treatment of the electric field will modify the dispersion operator, characterizing this change is the aim of that future study. Since the C-flow instability operates even for zero-mean $\mu$ (see \cref{fig:muSevol}), we expect it to proceed until the fluctuations of $\mu$ are completely depleted through the damping of Langmuir oscillations. We also expect the dynamical electric field to mitigate the clustering of $\mu$ seen during the runaway phase in \cref{fig:muSevol}. Without this dynamical treatment, the runaway reported here should be interpreted only qualitatively.

\sect{Summary}
We present a linear instability of magnetized plasmas, the charge-flow (C-flow) instability, driven by the transport of charge fluctuations by the bulk flow.
Its maximum growth rate for small $B_0$ at unit magnetic Prandtl
number, \(\abs{\omega_i}_{\rm max} = \cf \kk \abs{\mu} B_0/8\) is independent of the magnetic diffusivity even though finite diffusivity is required for the underlying current to enter the induction equation.
Direct numerical simulations confirm the growth rate to within a few percent with no free parameters (\cref{fig:spec}) and the predicted scaling with \(\cf\), \(B_0\), and \(\mu\) (\cref{fig:gamma}), and demonstrate that the instability operates for fluctuations of \(\mu\) with zero mean (\cref{fig:muSevol}), as required for a globally neutral plasma.
The driving current can exceed the chiral vortical current and become comparable to the chiral magnetic one, and is therefore relevant wherever chiral MHD is applied with \(\mu \neq 0\). The saturation of the instability, absent in the one-fluid approximation employed here, is governed by the longitudinal electric response, requiring a two-fluid
treatment that we defer to future work.

\sect{Data availability}
The run files and time-series data supporting this Letter are available on Zenodo~\cite{my:cflow_data}.

\sect{Acknowledgments}
We thank I. Rogachevskii for helpful discussions.

\appendix

\section{Derivation of the reduced model}

Let us consider the case of the early Universe, for which the equations already simplify significantly \cite[Eqs.~(61)-(69)]{my:chiralmhd_theory}, yielding the equations for the chemical potentials,
\begin{multline}
    \pd_\ttt \left[
    \mu
    + \frac{\cm}{\kk} \vv \cdot \qty(\del \times \bb')
    \right]
    + \del \cdot \left[ \vphantom{\frac{C_5}{k_5}}
    \mu \vec{v}
    + \cm \mu_5 \vec{B}'
    \right.\\\left.
    + \frac{\cm}{\eta \kk} \left(\vec{E}' +  \vec{v} \times \vec{B}' \right)
    + \frac{\cw}{\kk} \mu \mu_5 \vec{\omega}'
    \right] = 0 \,,
    \label{eq:mu0}
\end{multline}
\begin{multline}
    \pd_\ttt \mu_5 
    + \frac{\pi^2 \cw}{3 \kk} \vec{\omega} \cdot \pd_\ttt \vec{v}
    + \del \cdot \left[
    \mu_5 \vec{v}
    + \cm \mu \vec{B}'
    \right.\\\left.
    + \mu \mu_5 C_{\rm CESE} \left(\vec{E}' +  \vec{v} \times \vec{B}' \right)
    \right]
    + \frac{\cw}{2 \kk} \vec{\omega}' \cdot \del \left(\mu^2 + \mu_5^2\right)
    \\
    = -\frac{\pi}{2} \kk \qty(\frac{m_e}{T})^2 \mu_5 + 1.97 \mk \kk \mk \vec{E}' \cdot \vec{B}' \,,
    \label{eq:mu50}
\end{multline}
the equations for the electromagnetic fields
\begin{gather}
    \frac{\pd \vec{B}'}{\pd \ttt}
    = - \del \times \vec{E}'\,,
    \label{eq:ind0}
    \\[4pt]
    \ee' =  -\vec{v} \times \vec{B}'
    + \eta
    \left( \del \times \vec{B}'
    - \kk \mu_5 \vec{B}'
    - C_{\rm flow} \kk \mu \vec{v}
    \nonumber \right.\\\left.
    \mspace{263mu}
    - \frac{\cw}{\cm} \mu \mu_5 
    \vec{\omega}' \right) \,,
    \label{eq:elec0}
\end{gather}
and the various constants used
\begin{gather}
    \cm = \sqrt{\frac{3 \alpha}{5 \pi}}
    \simeq \num{3.7e-2} \,,
    \quad C_{\rm flow} = \frac{1}{\cm} \simeq \num{26.8} \,,
    \\
    \cw = \frac{6\alpha}{\pi^3} \simeq \num{1.4e-3} \,,
    \\
    C_{\rm CESE} = \frac{6 \cm}{\kk \eta \pi^2} \simeq \num{2.68e3} \,,
    \\
    \kk \equiv \frac{2 \alpha T_0}{\pi \hbar} \,,
\end{gather}
where all the dimensional quantities are scaled to their comoving values as described in \cite[Sec.~II]{my:chiralmhd_theory}, $\cm$ and $\eta$ are approximated for the early Universe as per \cite[Eqs.~(66)-(69)]{my:chiralmhd_theory}, $\vec{\omega}' = \vec{\omega} +  \vec{v} \times  \pd_\ttt \vec{v}$, $\vec \omega = \del \times \vv$ is the vorticity vector,  the magnetic field is normalized as $\bb' \equiv \bb/\sqrt{\mu_0\rho_0}$ and therefore measures the Alfv\'en speed associated with the field, the electric field is similarly normalized as $\ee' \equiv \ee/\sqrt{\mu_0\rho_0}$, \(\rho_0\) is the background plasma density, \(\mu_0\) is the vacuum permeability, $\ttt$ is the comoving time, temperature \(T = \min \{T_{\rm plasma}, \, T_{\rm EW}\}\) in \cref{eq:mu50} is the energy scale of the background plasma or the electroweak scale, whichever is less, $T_0$ is the present-day CMB temperature, $m_e$ is the mass of an electron, and $\alpha$ is the fine-structure constant. These equations are supplemented by the equations for hydrodynamics, which are unchanged by the direct chiral effects,
\begin{gather}
    {\frac{D \vec{v}}{D \ttt}}
    = \frac{\rho_0}{\rho} (\vec{\del} \times \vec{B}') \times \vec{B}'
    - \frac{\vec{\del} p}{\rho} + \frac{\vec{\del} \cdot (2 \nu \rho \vec{\msf{S}})}{\rho} \,,
    \label{eq:v00}
    \\[4pt]
    \frac{D\rho}{D \ttt} = - \rho \vec{\del}  \cdot \vec{v} \,,
    \label{eq:mass00}
\end{gather}
where $D/D\ttt \equiv \pd_\ttt + \vv\cdot\del$, $\vec{\msf S}$ is the rate-of-strain tensor, and $\nu$ is the kinematic viscosity.

In order to focus on the C-flow instability, the subject of this work, let us simplify these equations further. To that end, note that the chiral vortical effect and axial vortical effect terms, \ie the terms that are proportional to $\cw$, are smaller than the C-flow term if
\begin{gather}
    \cw \mu_5 \vec\omega' \ll \kk \vv \,.
    \label{eq:cve}
\end{gather}
Since $\cw \ll 1$ and in the case of chiral dynamo, the vorticity produced is at the scale of the order of $k_5 = \mu_5 \kk$ \cite{ref:brandenburg17,ref:schober18}, \cref{eq:cve} can be expected to be true in the small chemical potential limit \eqref{eq:smallmu}. Thus, we can neglect these terms.
As mentioned in \Cite{my:chiralmhd_theory}, the ohmic current term in \cref{eq:mu0}, \ie the third term in the second square bracket, assumes dissipation of charge imbalances through ohmic current, which is true only for over-damped plasma oscillations. In the early Universe, plasma oscillations can be underdamped \cite{ref:baym97,ref:lebellac96,ref:blaizot96,ref:arnold00}, which would imply that the charge imbalances might dissipate at a much slower timescale. This requires full treatment of the damping of the plasma oscillations while keeping the inertia of the electrons and positrons, which is deferred to future work. We detail this further within the discussion of saturation in the main text. For now, we drop this term and also assume that the chiral electric separation effect, the term with $C_{\rm CESE}$ as the prefactor, is negligible. We neglect the chirality-flipping term (first term on the right-hand side) in \cref{eq:mu50} since it is an additional effect pertaining to the chiral dynamo that remains an active area of research (see \Cites{tex:gurgenidze25,ref:skoutnev26,ref:boyarsky21,ref:boyarsky21b}). Finally, we assume that the second term in the time derivative in \cref{eq:mu0} is negligible compared to $\mu$ and we drop it. In addition, let us add hyperdiffusion in \cref{eq:mu0,eq:mu50} for numerical stability \cite{ref:schober22}. With all these terms dropped and hyperdiffusion added, we arrive at the reduced set of equations given by \crefrange{eq:mu}{eq:elec}, supplemented by \cref{eq:v00,eq:mass00} which remain unchanged.

\bibliography{chiral_plasma, dgarg_main}

\end{document}